\documentstyle[11pt,paspconf]{article}

\begin{document}

\title{Partial Coverage and Time Variability of 
Narrow-Line Intrinsic QSO Absorption Systems}

\author{T.A. Barlow and F. Hamann}
\affil{Center for Astrophysics \& Space Sciences, University of 
California-- San Diego, La Jolla, CA 92093-0424}

\author{W.L.W. Sargent}
\affil{Astronomy Department, California Institute of Technology, 
Pasadena, CA, 91125}

\begin{abstract}

It is possible to distinguish ``intrinsic'' absorption systems (gas
clouds within the quasar environment) from ``intervening'' systems
(gas clouds unrelated to the quasar phenomenon) by considering
certain observational properties.  We find that the most direct
determinations of the intrinsic nature of a system are time
variability and partial coverage of the background light source by
the absorption region.  Both of these conditions are unlikely to
occur in intervening systems.  We present Keck and HIRES data which
demonstrate both these phenomena.  We also summarize a list of
several other observational properties which appear to be indicative
of ``intrinsic'' systems.

\end{abstract}


\section{Introduction}

It is now clear that at least some of the narrow-line QSO absorption systems
are caused by gas related to the quasar environment.  We expect the majority
of these systems to have absorption redshifts ($z_a$) close to the systemic
redshift of the quasar (as determined from the quasar emission-line redshift,
$z_e$), although intrinsic narrow lines have been found at outflow velocities
ranging from near zero to more than 30,000 km s$^{-1}$.  ({\it e.g.} Hamann
et al. this volume.)

We can classify quasar absorption lines into three general categories:  (1)
``Intervening'' systems which are due to gas clouds unrelated to the quasar
that coincidentally fall along our line-of-sight to the quasar.  These may be
associated with the disks or haloes of galaxies or exist independent of
galaxies in intergalactic space.  (2) ``Intrinsic'' systems which are caused
by gas related to the quasar environment and probably ejected outward from
the quasar central engine.  And (3) systems which are due to gaseous regions
within the host galaxy of the quasar or within a galaxy cluster which
contains the quasar.  These latter absorbers may not be ejected from the
quasar, but are still affected by the quasar luminosity and environment.
Since such regions are within the gravitational well of the quasar galaxy,
there is a bias for these clouds to intercept our line-of-sight as compared
to intervening systems which are much more distant from the quasar.

The most obvious intervening systems are those with much smaller redshifts
than the quasar ($z_a<< z_e$) where emission from a galaxy has also been
detected near the same line-of-sight and is at the same redshift as the
absorber.  In contrast, the most obvious intrinsic systems are the so-called
broad absorption-lines (BALs) in BALQSOs, which have large velocity widths
($\ga$1,000 km s$^{-1}$) and always appear within 50,000 km s$^{-1}$ of the
emission line redshift.  However, we require additional criteria in order to
classify narrow absorption lines (less than a few hundred km s$^{-1}$) which
have no identified galactic conterpart.  Statistically, we know that most of
these narrow line systems are intervening because the systems are distributed
between $z$=0 and $z$=$z_e$ indicating that the majority of systems are
cosmologically distributed between us and the quasar ({\it e.g.} Steidel
1990.)  In this paper, we discuss criteria for distinguishing the intrinsic
narrow line systems from the intervening systems and present examples
demonstrating some of the observational properties of categories (2) and
(3).

\section{Intrinsic vs. Intervening Indicators}

Recent studies of metallicity, ionization, time variability, and
velocity structure 
of narrow $z_a\sim z_e$ systems have shown that certain
properties appear to distinguish narrow intrinsic systems from the
more ubiquitous intervening systems.  In order of how reliably they
indicate an intrinsic system, these properties include:
(1) time variability (Hamann et al.\ 1995),
(2) partial coverage of the background light source along our line-of-sight
by the absorption-line region
(cf. Wampler et al.\ 1995; Hamann et al.\ 1997; Barlow \& Sargent 1997),
(3) high electron density as derived from fine structure
lines as a distance indicator when combined with an ionization level
estimate (cf. Morris et al.\ 1986),
(4) spectropolarimetry revealing increased fractional
polarization in the lines relative to the continuum (so far only
observed in BALs, cf. Goodrich et al.\ 1995, Cohen et al.\ 1995),
(5) velocity structure, {\it e.g.} a ``correlated'' or ``smooth'' 
line profile shape across many components when observed
at spectral resolutions of $\la$10 km s$^{-1}$ FWHM
(in contrast, intervening systems often show several
apparently independent components),
(6) high metallicities ({\it e.g.} Petitjean et al.\ 1994),
(7) high ionization levels,
(8) proximity to BALs, and
(9) redshifts close to the emission line redshift
({\it e.g.} Anderson et al.\ 1987.)

We consider the first few properties as strong indicators of the intrinsic
nature of the system, and the last few properties as very weak indicators.
For example, we expect some intervening systems to occur close to the
emission-line redshift and close to BALs, to have high ionization lines, and
possibly to have high metallicities ($\sim$ solar.)  Conversely, we do not
expect all intrinsic systems to lie close to the emission redshift or close
to BALs.

Time variability is perhaps the most conclusive indicator.  If the intrinsic
absorber is photoionized by the quasar continuum source, and the quasar
source is known to vary, we expect the absorption line optical depths to also
vary.  (Intervening C~IV systems, on the other hand, are probably ionized by
a stable flux which permeates the intergalactic-medium.)  It is also possible
for intrinsic absorption to vary due to the motion of the region across our
line-of-sight.  The two variability effects can be distinguished by observing
two lines with different ionization levels.

Unfortunately, detecting time variability requires two observations separated
by $\ga$1 year during which the quasar happens to vary.  A more direct
approach (requiring only one epoch) is to determine the line-of-sight
coverage fraction ($C_f$) of the background light source by the absorption
region.  High signal-to-noise (S/N$\ga$30 per resolution element) and high
resolution ($\la$10 km s$^{-1}$ FWHM) spectra can be used to measure the {\it
apparent} $C_f$ of the absorbing clouds as determined by the residual
intensities of a resolved and separated doublet such as C~IV
$\lambda\lambda$1548,1550 (see Hamann et al. this volume.)

Direct density limits from excited-state fine structure lines are also
useful.  Intervening systems are expected to have low density ($\la$1 cm
$^{-3}$), while at least some intrinsic systems appear to have higher density
($\ga$1000 cm$^{-3}$) from time variability (Hamann et al. 1995.)
Unfortunately, the high ionization level of most intrinsic systems mean that
the most readily observed fine structure lines such as C~II$^*$ $\lambda$1335
generally are not observable.

Ionization and metallicity are subject to uncertainties in the
photoionization model and the shape of the ionizing continuum.  An
observational feature which may result from
both high ionization and high metallicity is the unusually high
N(N~V)/N(H~I) ratio (as compared to intervening systems) seen in many
intrinsic systems.

\section{Observations and Discussion}

We present HIRES data on two quasars with narrow $z_a\sim z_e$ absorption
which show partial coverage (Q0449-13, $z_e$=3.09) and time variability
(Q1700+64, $z_e$=2.72).  These systems also show high ionization lines,
possible high metallicity, and line profiles that are broader and less
structured than lines seen at high resolution in intervening
systems.  

\begin{figure}
\plotfiddle{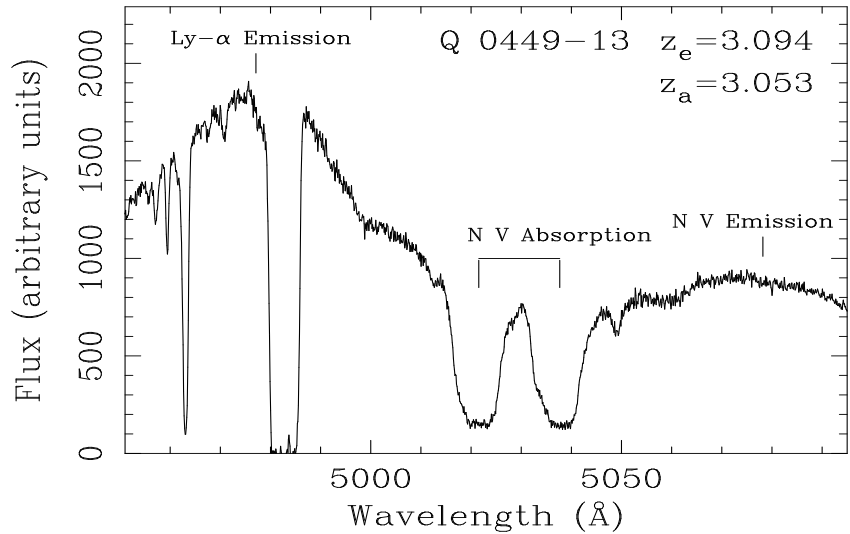}{2.5in}{-90.0}{48.}{40.}{-180.0}{+221.0}
\caption{Keck HIRES spectra of Q~0449--13 showing partial
coverage of the continuum $+$ emission flux by the 
N~V $\lambda\lambda$1238,1242 doublet.}
\end{figure}

Figure 1 shows the N~V $\lambda\lambda$1238,1242 doublet in the QSO
0449--13.  This system might be considered a ``mini-BAL'', since the width
(600 km s$^{-1}$) and outflow (3,000 km s$^{-1}$) are smaller than BALs in
other QSOs, but the smooth structure and width is significantly different
than intervening systems.  Note that both N~V lines appear to be saturated
but do not reach zero intensity.  This effect indicates that the absorption
region does not completely cover the background light source.  Since these
lines occur near the Ly$\alpha$ and N~V broad emission lines (BELs), it is
possible that the continuum is completely covered while the broad
emission-line region (BELR) is only partially occulted.  The ambiguity of BEL
and continuum coverage occurs in many $z_a$$\sim$$z_e$ systems, however by
studying various ions it is possible to distinguish the two effects (Barlow
\& Sargent 1997.)

\begin{figure}
\plotfiddle{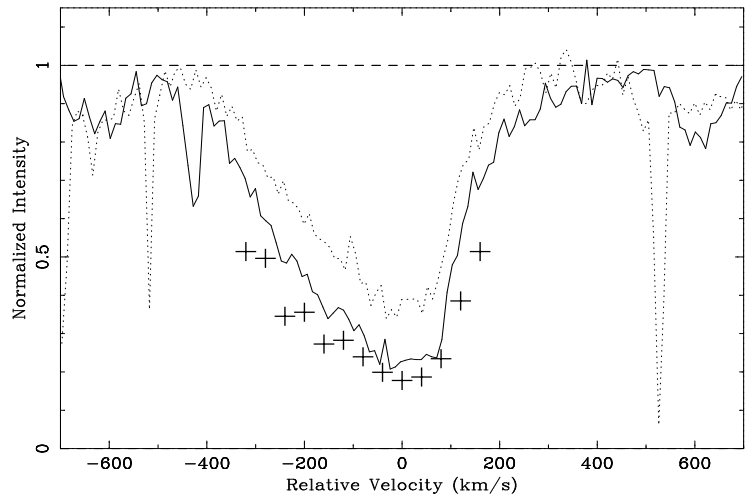}{2.5in}{-90.0}{40.}{37.}{-160.0}{+211.0}
\caption{Keck HIRES spectra of Q~0449--13 showing the 
Si~IV $\lambda\lambda$1393,1402 doublet
on a velocity scale (arbitrary zero-point.)
The crosses indicate the amount of background light
which is apparently unobscured by the absorption region.}
\end{figure}

Figure 2 shows both lines of the Si~IV doublet in Q~0449--13 on a velocity
scale.  The crosses indicate the amount of the light which must be subtracted
to obtain a 2:1 optical depth ratio.  In other words, these values indicate
the apparent fraction of the background light source not covered by the
absorption region.  The weakness of the Si~IV BEL means that at least some of
the continuum source is not obscured by the Si~IV absorption.  Note that
$C_f$ varies from $\sim$0.8 to $\sim$0.5 across the Si~IV line profile.  This
effect of a velocity dependent coverage fraction has been seen in at least
one BALQSO (CSO 755, cf. Barlow \& Junkkarinen 1994.)

Partial coverage by an absorption region creates a difficulty in determining
the true column density of a given ion.  In principle, it is possible to
determine the true optical depth if $C_f$ can be measured.  However, this
measurement is complicated by the fact that $C_f$ can vary between ions and
with velocity, and can depend upon the underlying BEL strength (Barlow \&
Sargent 1997.)

\begin{figure}
\plotfiddle{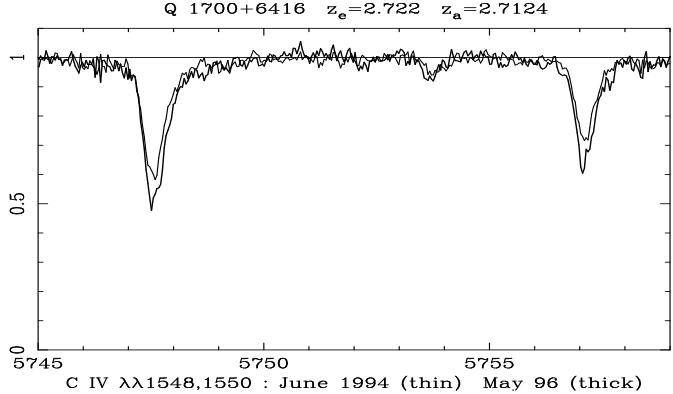}{1.8in}{-90.0}{37.}{27.}{-150.0}{+150.0}
\caption{Keck HIRES Spectra of Q~1700+64 showing time variability
and partial coverage by the C~IV $\lambda\lambda$1548,1550 doublet.
The emission-line plus continuum flux has been normalized.
The heliocentric, vacuum wavelengths are in Angstroms.}
\end{figure}

Figure 3, shows time variability in the C~IV doublet for the $z_a\sim z_e$
system in 1700+64 for two epochs separated by two years.  The outflow
velocity, using $z_e$=2.722, is 770 km s$^{-1}$.  The apparent optical depth
of the C~IV lines decreased by about 25\%.  The N~V doublet also varied by a
similar amount during the same period.  This system yields $C_f<1$ in both
C~IV and N~V.  The line width (FWHM$\sim$30 km s$^{-1}$) is well within the
range of intervening systems, but the gas is clearly intrinsic to the
quasar.

Variability similar to the case in Q1700+64 can be detected in $\sim$30\% of
intrinsic narrow-line systems with data of comparable resolution and S/N
(Barlow et al.\ 1997, in preparation.)  Variations on a $\sim$1 year
timescale indicate electron densities $\ga$1000 cm$^{-3}$, since the
recombination time ($\propto$ 1/$n_e$) must be short enough for the
ionization levels to reach equilibrium.

\begin{figure}
\plotfiddle{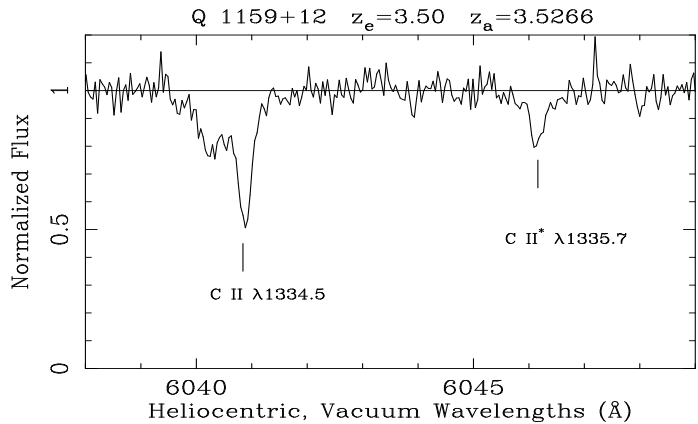}{2.0in}{-90.0}{38.}{32.}{-172.0}{+172.0}
\caption{Keck HIRES Spectra of Q~1159+12 showing the fine structure
line C~II $\lambda$1335.}
\end{figure}

Figure 4 shows Keck HIRES data of the QSO 1159+12 ($z_e$=3.50) which contains
an absorption system at $z_a=3.5266$ with a velocity width of 60 km s$^{-1}$
and an apparent {\it infall} velocity of about 1700 km $^{-1}$ relative to
the C~IV BEL.  This system includes the low-ionization line C~II
$\lambda$1334 and the fine structure line C~II$^*$ $\lambda$1335, as well as
C~IV and N~V.  From the ratio N(C~II)/N(C~II$^*$), we compute an electron
density of about 6 cm$^{-3}$.  By estimating an ionization parameter of
$\approx$0.06 from Si~II, Si~III, and Si~IV, we obtain a distance from the
quasar of 800 kpc.  Although this system has $z_a> z_e$ and high ionization
lines, it is clearly outside the near quasar environment and not included in
the class of ejected intrinsic absorbers.  However, since it is ionized by
the quasar and may reside in the galaxy cluster of the quasar, we place this
system in the third category of absorber listed in the introduction.

With the aid of high resolution, high signal-to-noise data, narrow intrinsic
absorption systems can be distinguished from intervening systems unrelated to
quasar.  By classifying these systems and studying their correlated
properties such as density and metallicity, we can study the evolution of the
quasar environment and determine the effects on the quasar host galaxy and
galaxy cluster.

\acknowledgments

This work has been supported by grants AST92-21365 and AST95-29073
from the National Science Foundation.  This work has also been supported
in part by NASA grant NAS5-26555.

\end{document}